\documentclass[prb,showpacs,amsmath,amssymb]{revtex4}

\usepackage{amsmath,amsthm,amssymb,amstext}
\usepackage{graphicx}
\usepackage{graphics}
\usepackage{citesort}
\usepackage{enumerate}
\usepackage{array}

\newtheorem{prop}{Proposition}
\newtheorem{lemma}{Lemma}

\newcommand{\ai}{{\rm Ai}}
\newcommand{\aip}{{\rm Ai}\,'}

\newcommand{\gue}{\text{GUE}}
\newcommand{\lue}{\text{LUE}}
\newcommand{\jedge}{\tilde{J}_N(\xi)}
\newcommand{\re}{\text{Re}}
\newcommand{\im}{\text{Im}}

\newcommand{\Arcsin}{{\rm Arcsin}}
\newcommand{\Arccos}{{\rm Arccos}}
\newcommand{\asym}{\mathop{\sim}\limits^{.}}
\newcommand{\rhosoft}{\frac{(bN)^{1/3}}{2}\rho_N\left(1+\frac{b^{1/3}}{2}\frac{\xi}{N^{2/3}}\right)}

\begin{document}

\title{Asymptotic corrections to the eigenvalue density of the GUE and LUE}

\author{T. M. Garoni}
\affiliation{Institute for Mathematics and its Applications, 
University of Minnesota, 400 Lind Hall, 207 Church Street S.E.,
Minneapolis, MN 55455-0436, USA}
\email{garoni@ima.umn.edu}
\homepage{http://www.ima.umn.edu/~garoni}

\author{P. J. Forrester}
\affiliation{Department of Mathematics and Statistics, University of
  Melbourne, Parkville, Victoria 3010, Australia}
\email{P.Forrester@ms.unimelb.edu.au}
\homepage{http://www.ms.unimelb.edu.au/~matpjf/matpjf.html}

\author{N. E. Frankel}
\affiliation{School of Physics, University of
  Melbourne, Parkville, Victoria 3010, Australia}
\email{n.frankel@physics.unimelb.edu.au}

\date{\today}

\begin{abstract}
We obtain correction terms to the large $N$ asymptotic expansions of
the eigenvalue density for the Gaussian unitary and Laguerre unitary ensembles of random
$N\times N$ matrices, both in the bulk of the spectrum and near the
spectral edge. This is achieved by using the well known
orthogonal polynomial expression for the kernel to construct a double
contour integral representation for the density, to which we apply the
saddle point method. The main correction
to the bulk density is oscillatory in $N$ and depends on the 
distribution function of the limiting density, while the corrections
to the Airy kernel at the soft edge are again expressed in terms of
the Airy function and its first derivative. We demonstrate numerically
that these expansions are very accurate.  A matching is exhibited between
the asymptotic expansion of the bulk density, expanded about the edge,
and the asymptotic expansion of the edge density, expanded into the bulk.
\end{abstract}

\maketitle

\section{Introduction}
We consider in this paper two classical ensembles of random matrices, the Gaussian
unitary ensemble (GUE), 
and the Laguerre unitary ensemble (LUE).
These ensembles can be characterized by their
joint eigenvalue probability density functions
\begin{equation}
P_N(x_1,\dots,x_N)\propto
\prod_{l=1}^N \omega_N(x_l)\,\prod_{1\le j<k\le N}(x_k-x_j)^2,\qquad
x_l \in \Omega,
\label{eigenvalue density}
\end{equation}
with
\begin{equation}
\omega_N(x)=
\begin{cases}
\exp(-2N x^2), & \gue,\\
x^{\alpha}\,\exp(-4N x), & \lue,\\
\end{cases}
\label{weight definition}
\end{equation}
and
\begin{equation}
\Omega
=
\begin{cases}
\mathbb{R},& \gue,\\
(0,\infty),& \lue.\\
\end{cases}
\end{equation}
The GUE consists of $N\times N$ Hermitian matrices with independent
normally distributed entries on and above the diagonal. It is the
cornerstone of random matrix theory \cite{ForresterBook,Mehta,Deift}. The 
LUE has fundamental applications in 
mathematical statistics and quantum field theory since it 
includes Wishart matrices and the Chiral GUE as special cases (the latter after a straightforward
change of variables); see e.g. Ref. \cite{ForresterBook}.

We are interested in the large $N$ behavior of the marginal
eigenvalue probability density $\rho_N(x)$, which 
we hereafter refer to simply as ``the density'', and which is defined by
\begin{equation}
\rho_N(x):=\int_{\Omega^{N-1}}P_{N}(x,x_2,\dots,x_N)\,dx_2\dots dx_N.
\label{density definition}
\end{equation}
The function $N\,\rho_N(x)$ can be interpreted as the 
number density of eigenvalues near the point $x$. 
We also remark that for the GUE, $N\,\rho_N(x)$ is equal to the number density
of a harmonically trapped system of either non-interacting fermions or
impenetrable bosons \cite{ForresterFrankelGaroniWitte03a}. 
There is a
similar interpretation for the LUE in terms of a Calogero-Sutherland
type model \cite{ForresterFrankel04}. For recent advances in asymptotic questions
related to these interpretations, complementary to the present study, see
Refs. \cite{ForresterFrankelGaroniWitte03b,Garoni05,Krasovsky04}

As background to the present study we note that aspects of the large $N$ form of
$\rho_N(x)$ first arose in studies of field theories related to Hermitian matrix
models \cite{BrezinItzyksonParisiZuber78}. There, for the GUE the large $N$ asymptotic expansion of the
moments
$$
m_N(p) := \int_\Omega x^p \rho_N(x) \, dx \qquad (p=1,2,\dots)
$$
was sort. By a graphical expansion of the matrix integral, involving cataloging
the corresponding maps according to their genus, it was predicted that for certain
coefficients $a_{2j}(p)$,
\begin{equation}
m_N^{\rm GUE}(p) = \sum_{j=0}^{p/2} \frac{a_{2j}(p)}{N^{2j} }, \qquad (p=2,4,\dots)
\label{GUE moments}
\end{equation}
(the odd moments of course vanish). Analogous considerations in the case of the LUE 
\cite{DiFrancesco02} show that
\begin{equation}
m_N^{\rm LUE}(p) = \sum_{j=0}^{p/2} \frac{\tilde{a}_{j}(p,\alpha)}{N^{j} } 
\label{LUE moments}
\end{equation}
for certain coefficients $\tilde{a}_j(p,\alpha)$. Observe in particular that (\ref{GUE moments}) contains
only even inverse powers in $N$, while (\ref{LUE moments}) contains both even and odd inverse powers in $N$.

The graphical methods allow $a_0(p)$ in (\ref{GUE moments}) and $\tilde{a}_0(p,\alpha)$ in (\ref{LUE moments}) to
be computed in terms of binomial coefficients for all $p=0,1,\dots$. This knowledge in turn
can used (see e.g.~\cite{ForresterBook}) to prove that in the limit $N \to \infty$
and with $x$ fixed 
\begin{equation}
\rho(x):=\lim_{N\to \infty}\,\rho_N(x)=
\begin{cases}
\displaystyle
\frac{2}{\pi}\sqrt{1-x^2}\quad,x\in[-1,1],&\gue,\\
\displaystyle
\frac{2}{\pi}\sqrt{\frac{1}{x}-1}\quad,x\in(0,1],&\lue,\\
0,& otherwise.\\
\end{cases}
\label{bulk limit}
\end{equation}
The first functional form in (\ref{bulk limit}) is referred to as the Wigner semi-circle law, while the second is
sometimes named after
Mar\u{c}enko-Pastur.
See e.g. Refs. \cite{Mehta,ForresterBook,Deift}.  

The expansions (\ref{GUE moments}) and (\ref{LUE moments}) provide a motivation to undertake a study of the
asymptotic form of (\ref{density definition}). In the case of the GUE such a result has been
given by Kalish and Braak \cite{KalischBraak}. It states that for $|x| < 1$ and fixed
\begin{equation}
\rho_N(x)= \rho(x)-\frac{2\cos[2\,N\,\pi\,P(x)]}{\pi^3\, \rho^2(x)}\frac{1}{N}
+O\left(\frac{1}{N^2}\right),
\label{Kalish Braak}
\end{equation}
where 
$$
P(x) = 1 + \frac{x}{2} \rho(x) - \frac{1}{ \pi} {\rm Arccos}(x).
$$
Thus one sees that unlike the situation with the moments (\ref{GUE moments}), the leading correction
term is $O(1/N)$. Of course this term is oscillatory so one might anticipate that after integration
it contributes at a higher order. However inspection of (\ref{Kalish Braak}) reveals that the
situation is more complex: the oscillatory term is not integrable at the endpoints of the support
$|x|=1$. Indeed, it is well known that with the boundary of the eigenvalue support taken as the
origin, a scaling regime distinct from that of the bulk becomes relevant. Explicitly, with
$\ai(x)$ denoting the Airy function, it has been proved that \cite{Forrester93}
\begin{equation}
\begin{split}
\lim_{N\to \infty}\,\frac{N^{1/3}}{2}\,\rho_N^{\rm GUE}\left(1+\frac{\xi}{2 N^{2/3}}\right)
&=
\lim_{N\to \infty}\frac{(2N)^{1/3}}{2}\,\rho_N^{\rm LUE}\left(1+\frac{\xi}{(2N)^{2/3}}\right)
\\&=[\ai'(\xi)]^2-\xi [\ai(\xi)]^2,
\label{soft edge limit}
\end{split}
\end{equation}
where $\xi$ is fixed. In view of the breakdown of (\ref{Kalish Braak}) in the vicinity of the
spectrum edge, (referred to as the {\it soft edge}, since although it defines the edge of 
the support of $\rho(x)$, for any finite $N$ there is a nonzero 
probability of finding eigenvalues lying beyond it),
we are thus led to also investigate the large $N$ asymptotic expansion extending
the limit law (\ref{soft edge limit}).

At a technical level, the main achievement of this paper is the derivation of the first
correction terms to the limit laws   (\ref{bulk limit}) and  (\ref{soft edge limit}).
We do this by utilizing the well known
orthogonal polynomial expression for $\rho_N(x)$ to obtain a double
integral representation which is amenable to the saddle point method.
In the bulk, i.e. in the interior of the support of $\rho(x)$, 
we show that the asymptotic series progresses in powers
of $1/N$, and we obtain the explicit form of the $1/N$ correction. We
find for the LUE that the coefficient of the $1/N$ term 
consists of a component which is oscillatory in $N$ as well a
component which is non-oscillatory in $N$, 
whereas for the GUE it consists of only an oscillatory component, as
shown in (\ref{Kalish Braak}). For
the soft edge we will see that the asymptotic series progresses in
powers of $N^{-1/3}$, and we obtain explicit expressions for the
coefficients of the $N^{-1/3}$ and $N^{-2/3}$ terms, which again involve Airy functions. 

Due to the similarity in the structure of $\rho_N(x)$ for
the GUE and LUE, it is convenient to consider both cases
simultaneously to avoid unnecessary repetition, and so at each step of
our presentation we discuss the GUE and LUE in parallel. 
In Section \ref{contour integrals} we discuss the double contour
integral expression for $\rho_N(x)$ to which we shall apply the saddle
point method. Section \ref{bulk} contains our discussion of the
asymptotics of $\rho_N(x)$ in the bulk while Section \ref{soft edge}
discusses the soft edge. In Section \ref{matching} we discuss the
extent to which our expansions in the bulk match up with those for the
soft edge.

\section{Contour integral expression for $\rho_N(x)$}
\label{contour integrals}
For the unitary ensembles there is a well known and very neat expression for
$\rho_N(x)$ in terms of
orthogonal polynomials, valid for any $N$ and $x\in\Omega$. If we let
$\{\pi_j(x)\}_{j=0}^{\infty}$ denote the monic polynomials orthogonal
with respect to $\omega_N(x)$ on $\Omega$, then 
\begin{equation}
\rho_N(x)=\frac{\omega_N(x)}{N\,\|\pi_{N-1}\|^2}
[\pi_N'(x)\,\pi_{N-1}(x)-\pi_{N-1}'(x)\,\pi_N(x)
].
\label{Christoffel-Darboux}
\end{equation}
The norm in the denominator of (\ref{Christoffel-Darboux}) is just the
$L^2$-norm associated with $\omega_N(x)$ and $\Omega$.
For a derivation of (\ref{Christoffel-Darboux}) the reader is referred
to Refs. \cite{ForresterBook,Mehta,Deift}. We remark at this point that there is no
universally agreed scale by which the GUE and LUE are defined. To match our choice of
scale and notation in (\ref{weight definition}) to that employed in
Ref. \cite{ForresterBook} for instance, we observe that 
$$
\rho_N(x)= 
\begin{cases}
\sqrt{2}N^{-1/2}\, P_N(\sqrt{2N} x,\sqrt{2N}x),& \gue,\\
4\, P_N(4N x,4N x),& \lue,\\
\end{cases}
$$
where $P_N(x,y)$ (not to be confused with our definition 
(\ref{eigenvalue density})
above) is the kernel
defined in Chapter 4 of Ref. \cite{ForresterBook} (the kernel is often also denoted $K_N(x,y)$ in the
literature) .

To investigate the large $N$ behavior of $\rho_N(x)$ it is obviously
advantageous to start with the expression (\ref{Christoffel-Darboux})
rather than with the $(N-1)$-fold integral (\ref{density definition}).  
The $\pi_{N+j-1}(x)$ can be expressed in terms of the standard Hermite and
Laguerre polynomials found in  Szeg\"o's classic book\cite{Szego} as follows
\begin{equation}
\pi_{N+j-1}(x)=
\begin{cases}
2^{-3(N+j-1)/2}N^{-(N+j-1)/2}H_{N+j-1}(\sqrt{2N}x), & \gue,
\\
(-1)^{N+j-1}(N+j-1)!(4N)^{-N-j+1}\, L_{N+j-1}^{(\alpha)}(4Nx), & \lue.
\\
\end{cases}
\label{mops as scaled classic ops}
\end{equation}
The required asymptotic expansions of the scaled Hermite and Laguerre
polynomials appearing in (\ref{mops as scaled classic ops}) are known to any order
both in the bulk and near the soft edge \cite{PlancherelRotach,Moecklin}, and such
asymptotic expansions of scaled orthogonal polynomials are now generically said to be of {\em
  Plancherel-Rotach} type (Plancherel and Rotach were the first
to compute such asymptotics for the Hermite polynomials). It is
reasonable to assume that
the most straightforward procedure to obtain the desired asymptotic
corrections for $\rho_N(x)$ in each region of interest is to simply
insert the corresponding asymptotic expansion for $\pi_{N+j-1}(x)$ into
(\ref{Christoffel-Darboux}). While this is certainly legitimate in
principle, and does indeed recover the leading term fairly easily, to
derive the correction terms it
turns out that such a procedure is rather tedious, and provides little if any insight into the resulting
expressions. The source of the complication is that the asymptotic
expansions for $\pi_{N+j-1}(x)$ contain a large amount of superfluous
information which is canceled when the expansions are substituted
into (\ref{Christoffel-Darboux}). To avoid this, we shall pursue a related, but more direct route.

The Plancherel-Rotach asymptotics for the Hermite and Laguerre polynomials 
were originally derived by first expressing the polynomials in terms of
contour integrals, and then applying the saddle point method.
By suitably massaging the 
standard results in Szeg\"o's book \cite{Szego} one finds that
\begin{align}
\pi_{N+j-1}(x)
&=
\begin{cases}
\begin{displaystyle}
c_j(N)
\,
\oint \frac{dz}{2\pi i}\,
e^{2N\, z\,x}\frac{e^{- N z^2/2}}
{z^{N+j}},
\end{displaystyle}
& 
\gue,\\
\begin{displaystyle}
(-1)^{N+j-1}c_j(N)
 \oint \frac{dz}{2\pi i}\,
e^{-2N\, z\,x}\frac{(z+2)^{N+\alpha}}{z^{N+1}}\left(\frac{1}{z}+\frac{1}{2}\right)^{j-1},
\end{displaystyle}
&
\lue,\\
\end{cases}
\label{pi integrals}
\\
c_j(N)
&:= 
\frac{(N+j-1)!}{(2N)^{N+j-1}}.
\end{align}
In both cases the contour of integration is a closed positively oriented contour 
which encircles the origin; in the Laguerre case we further demand that it not contain the point 
$z=-2$.

Instead of applying the saddle point method to (\ref{pi integrals}) and
then substituting the expansions into (\ref{Christoffel-Darboux}),
we shall first insert the contour integrals (\ref{pi integrals}) into
(\ref{Christoffel-Darboux}) to obtain a double integral expression for
$\rho_N(x)$, and then perform the saddle point method on this double
integral. 
To highlight the similarity between the GUE and LUE it is
convenient in the GUE case to substitute the contour integral for $\pi_{N+j-1}(-x)$
into (\ref{Christoffel-Darboux}) rather than that for $\pi_{N+j-1}(x)$;
since $H_{j}(-x)=(-1)^j\,H_j(x)$ this ruse is perfectly harmless.
This results in
\begin{equation}
\rho_N(x)=2 \frac{c_0(N)\,c_1(N)}{\|\pi_{N-1}\|^2}\omega_N(x)
J_N(x),
\label{density double integral}
\end{equation}
where 
\begin{equation}
J_N(x):=
\oint\frac{dz_1}{2\pi i}\oint\frac{dz_2}{2\pi i} e^{N S(z_1,x)+N S(z_2,x)}
G(z_1,z_2)
\label{J definition}
\end{equation} 
and
\begin{align}
S(z,x)&:=
\begin{cases}
-2z\,x -\log(z)-z^2/2,& \gue,\\
-2z\,x -\log(z)+\log(1+z/2), &\lue,\\
\end{cases}
\label{action definition}
\\
G(z_1,z_2)&:=u(z_1)\,u(z_2)\,\left(1-\frac{z_1}{z_2}\right),
\label{G definition}
\\
u(z)&:=
\begin{cases}
1,& \gue,\\
(1+z/2)^{\alpha-1},& \lue.\\
\end{cases}
\end{align}
The remainder of this paper will involve a careful asymptotic analysis
of the double integral (\ref{J definition}).

Before proceeding we note that it is straightforward to show, using standard results in the
orthogonal polynomial literature \cite{Szego}, that 
\begin{equation}
\|\pi_{N-1}\|^{-2}
=
\begin{cases}
\displaystyle
\frac{2^{2N-3/2}}{\sqrt{\pi}}\,\frac{N^{N+1/2}}{N!},
& \gue,\\
\displaystyle
\frac{(4N)^{2N+\alpha-1}}{\Gamma(N) \,\Gamma(N+\alpha)},
& \lue,\\
\end{cases}
\end{equation}
and hence the asymptotics of the prefactors in (\ref{density double integral}) is 

\begin{align}
2 \frac{c_0(N)\,c_1(N)}{\|\pi_{N-1}\|^2}
&=
\begin{cases}
\displaystyle
{\sqrt{\frac{2}{\pi }}}\,N^{\frac{3}{2} - N}\,\Gamma(N),&\gue,\\
\displaystyle
\frac{4^{N + \alpha }\,N^{\alpha }\,\Gamma(1 + N)}{\Gamma(N + \alpha )},&\lue,\\
\end{cases}
\\
&=
\begin{cases}
\displaystyle
2N e^{-N} \left[1  + \frac{1}{12\,N}
% + \frac{1}{288\,N^2}
+O\left(\frac{1}{N^2}\right)\right],&\gue,\\
\displaystyle
4^{N + \alpha }\,N
\left[1 -\frac{\left(\alpha-1\right) \,\alpha }{2\,N} 
%+ \frac{\left( -1 + \alpha  \right) \,\alpha \,\left( 1 + \alpha  \right) \,\left( -2 + 3\,\alpha  \right) }   {24\,N^2}
+O\left(\frac{1}{N^2}\right)
\right],&\lue.\\
\end{cases}
\label{prefactor asymptotics}
\end{align}

\subsection*{Saddle points}
%\label{saddle points section} 
Before applying the saddle point method to (\ref{J definition}) we need to
identify and classify the saddle points of (\ref{action definition}). 
The functions $S(z,x)$ in general have two saddle points at $z=z_{\pm}$ where
\begin{equation}
z_{\pm}:= 
\begin{cases}
-x \pm i \,\nu(x),&\gue,\\
-1\pm i \,\nu(x), &\lue,\\
\end{cases}
\label{saddle points}
\end{equation}
and
\begin{equation}
\nu(x):=
\begin{cases}
 \sqrt{1-x^2},&\gue,\\
\displaystyle 
\sqrt{\frac{1}{x}-1},&\lue.\\
\end{cases}
\label{nu definition}
\end{equation}
We note that for both the GUE and LUE we have 
\begin{equation}
\nu(x)=\frac{\pi}{2}\rho(x), \quad \text{for}\quad |x|\le 1,
\end{equation}
with $\rho(x)$ as defined in (\ref{bulk limit}).

Since for $|x|\le 1$ we have
\begin{equation}
\frac{S''(z_{\pm},x)}{2}=
\begin{cases}
\nu(x)\, e^{\pm i(\pi-\Arcsin(x))},&\gue,\\
2x^2\,\nu(x)\,e^{\pm i \pi/2},&\lue,\\
\end{cases}\label{bulk second S derivative}
\end{equation}
the saddle points $z=z_{\pm}$ are both simple when $|x|<1$, i.e. 
$S''(z_{\pm},x)\not = 0$.
However, when $x=1$ the two simple saddle points given in 
(\ref{saddle points})
coalesce to $z_{\pm}=-1$ and $S''(z_{\pm},1)$ vanishes, 
so we obtain one double saddle point in this case. Thus we already see why the
regions $|x|<1$ and $x\sim 1$ have qualitatively distinct asymptotic
behavior. Simple saddle points generically produce Gaussian integrals
whereas double saddle points generically produce Airy functions (see
e.g. \cite{Wong}).

\section{Bulk asymptotics for the GUE and LUE}
\label{bulk}
In the bulk of the spectrum, i.e. for $|x|<1$, we hold $x$ fixed and investigate
the asymptotics of (\ref{J definition}) as $N$ becomes large.
From (\ref{saddle points}) we see that there are two distinct simple saddle points of $S(z,x)$, which form a
complex conjugate pair in this case. Let's define
$S_{\pm}:=S(z_{\pm},x)$. Then since 
\begin{equation}
\re(S_+) =\re(S_-),
\end{equation}
both saddle points contribute to the same order and we deform our
contour through both of them. Denoting by $\Omega_{\pm}$
a contour passing through $z_{\pm}$ along a path of
steepest descent, the standard arguments of the saddle point method
(see for e.g. Ref. \cite{Wong}) yield
\begin{align}
J_N(x) &= 
\left(\int_{\Omega_+}\frac{dz_1}{2 \pi i}+\int_{\Omega_-}\frac{dz_1}{2 \pi i}\right)
\left(\int_{\Omega_+}\frac{dz_2}{2 \pi i}+\int_{\Omega_-}\frac{dz_2}{2 \pi i}\right)
e^{N\,S(z_1,x)+N\,S(z_2,x)}\,G(z_1,z_2)
+ O(e^{2N\,\re(S_+)-N\,\epsilon}),
\\
&=-\sum_{\sigma\in\{+,-\}^2}
\int_{\Omega_{\sigma_1}}\frac{dz_1}{2 \pi}
\int_{\Omega_{\sigma_2}}\frac{dz_2}{2 \pi}
\,e^{N\,S(z_1,x)+N\,S(z_2,x)}\,G(z_1,z_2)
+O(e^{2N\,\re(S_+)-N\,\epsilon}),
\label{bulk J first step}
\end{align}
for suitably small $\epsilon>0$.

We now need to parameterize the contours $\Omega_{\pm}$.
From (\ref{bulk second S derivative})
we see that if we set
\begin{equation}
\theta=
\begin{cases}
\Arcsin(x)/2,&\gue,\\
\pi/4,&\lue,\\
\end{cases}
\label{bulk steepest descent angles}
\end{equation}
then a suitable parameterization of $\Omega_{\pm}$ is
\begin{equation}
z=z_{\pm}+e^{\pm i \theta}t,\qquad t\in[-\eta,\eta],
\label{bulk parameterization}
\end{equation}
for sufficiently small $\eta>0$.
In (\ref{bulk steepest descent angles}) 
the function $\Arcsin:[-1,1]\to[-\pi/2,\pi/2]$ denotes the principle branch of arcsine. 
With the parameterization (\ref{bulk parameterization}) the contour
$\Omega_+$ is traversed in the negative direction, so we need to compensate for this
with an explicit minus sign. 

By choosing $\eta$ sufficiently small $S(z,x)$ is analytic on
$\Omega_{\pm}$ and so using the parameterization (\ref{bulk parameterization})
we see that
\begin{equation}
S(z,x)=
S_{\pm}- a\,t^2-a\,t^2\varphi_{\pm}(t),\qquad z\in\Omega_{\pm},
\label{bulk S series}
\end{equation}
where
\begin{equation}
a:=
\begin{cases}
\nu(x),&\gue,\\
2x^2\nu(x),&\lue,\\
\end{cases}
\end{equation}
and
\begin{equation}
\varphi_{\pm}(t):=\sum_{k=3}^{\infty}\frac{S^{(k)}(z_{\pm},x)}{S^{(2)}(z_{\pm},x)}\frac{2}{k!} 
e^{\pm i (k-2)\theta}\,t^{k-2}.
\end{equation}
We note that $\varphi_{-}(t)=\varphi_+^*(t)$, where $*$ denotes complex
conjugation.

It is a straightforward exercise to show that $S_-=S_+^*$, and that
\begin{align}
\re(S_+)
&=
\begin{cases}
\displaystyle
\frac{1}{2}+x^2,&\gue,\\
2x-\log(2),&\lue,\\
\end{cases}
\label{Re S}
\\
\im (S_+)
&=-\pi\, P(x),
\label{Im S}
\end{align}
where 
\begin{align}
P(x)&:=\int_{x_0}^x\,\rho(t)\,dt,
\\
&\phantom{:}=
\begin{cases}
\displaystyle
1+\frac{x}{2}\,\rho(x)-\frac{1}{\pi}\Arccos(x),&\gue,\\
\displaystyle
1+x\,\rho(x)-\frac{2}{\pi}\Arccos(\sqrt{x}),&\lue.\\
\end{cases}
\label{P definition}
\end{align}
Here $x_0$ is the left
edge of the support of $\rho(x)$ given in (\ref{bulk limit}), i.e. $x_0=0,1$, for the LUE and GUE respectively.
We note that $P(x)$ is the probability distribution
function corresponding to $\rho(x)$.
The limiting distribution function $P(x)$ will play a
significant role in the bulk asymptotic expansion of $\rho_N(x)$.

With the results (\ref{bulk S series}) and (\ref{Im S}) for $S(z,x)$, and the definitions 
\begin{align}
G_{\sigma}(t)
&:=
G(z_1,z_2)
\Big\vert_{\genfrac{}{}{0pt}{}{z_1\to z_{\sigma_1}+e^{i\sigma_1 \theta}t_1}{z_2\to z_{\sigma_2}
+e^{i\sigma_2 \theta}t_2}},
\label{bulk G transform}
\\
\mathbb{E}^{(B)}_i
&:=\int_{-\eta}^{\eta}\frac{e^{-a N t_i^2}}{2\pi}\,dt_i,
\label{bulk integral operator}
\end{align}
we can make the change of variables (\ref{bulk parameterization}) in 
(\ref{bulk J first step}) to obtain 
\begin{equation}
\begin{split}
J_N(x)
&=
e^{2N\,\re(S_+)}\,
\mathbb{E}^{(B)}_1\mathbb{E}^{(B)}_2\sum_{\sigma\in\{+1,-1\}^2}
-\sigma_1\sigma_2\,e^{i(\sigma_1+\sigma_2)(\theta-N\,\pi\,P(x))}
e^{-a N t_1^2 \varphi_{\sigma_1}(t_1) -a N t_2^2 \varphi_{\sigma_2}(t_2)}
G_{\sigma}(t)
\\
\quad&+O(e^{2N\,\re(S_+)-\epsilon\,N}).
\label{bulk parameterized}
\end{split}
\end{equation}
A small amount of massaging shows that
the $-,+$ term in (\ref{bulk parameterized}) is the complex conjugate of the $+,-$ term, and
likewise the $-,-$ term is conjugate to the $+,+$ term. 
Suppose now that we define the function $F_{\sigma}(\lambda,t)$ by
\begin{equation}
F_{\sigma}(\lambda,t):=
e^{\lambda_1\varphi_{\sigma_1}(t_1)}
e^{\lambda_2\varphi_{\sigma_2}(t_2)}
G_{\sigma}(t)
\label{bulk F definition}
\end{equation}
then
\begin{equation}
\begin{split}
J_N(x)
&=
2\,e^{2N\,\re(S_+)}
\,\re \,\mathbb{E}^{(B)}_1\mathbb{E}^{(B)}_2 \left\{\, F_{+,-}(\lambda,t)\Big\vert_{\lambda_i=-aNt_i^2}
-
e^{2i(\theta-N\,\pi\, P(x))}\,
 F_{+,+}(\lambda,t)\Big\vert_{\lambda_i=-aNt_i^2}\right\}
% \\
% &\quad
+O(e^{2N\,\re(S_+)-\epsilon N}).
\label{bulk J in terms of F}
\end{split}
\end{equation}
We shall discuss the purpose of the parameters $\lambda_i$ shortly.
Note that the two terms in (\ref{bulk J in terms of F}) are
qualitatively distinct \--- the second term is oscillatory in $N$
while the first is not.

It is convenient to pause for a moment and multiply 
(\ref{bulk J in terms of F}) by the 
the explicit forms for the prefactors required in (\ref{density double
integral}) using (\ref{prefactor asymptotics}) and (\ref{Re S}) to
obtain the corresponding expression for $\rho_N(x)$. Since
\begin{equation}
\frac{2c_0(N)\,c_1(N)}{\|\pi_{N-1}\|^2}\omega_N(x)\,e^{2N\,\re(S_+)}=
h_0\,N\left[1+\frac{h_1}{N}+O\left(\frac{1}{N^2}\right)\right]
\end{equation}
with
\begin{align}
h_0&=
\begin{cases}
2,&\gue,\\
4^{\alpha}\,x^{\alpha},&\lue,\\
\end{cases}
\\
h_1&=
\begin{cases}
\displaystyle
\frac{1}{12},&\gue,\\
\displaystyle
-\frac{\alpha(\alpha-1)}{2},&\lue,\\
\end{cases}
\end{align}
we have
\begin{equation}
\begin{split}
\rho_N(x)&=h_0\left[1+\frac{h_1}{N}+O\left(\frac{1}{N^2}\right)\right]
%\\&\quad\times
2\re\,N\,\mathbb{E}^{(B)}_1\mathbb{E}^{(B)}_2 \left\{F_{+,-}(\lambda,t)\Big\vert_{\lambda_i=-aNt_i^2}
  -
e^{2i(\theta-N\,\pi\, P(x))}\,
 F_{+,+}(\lambda,t)\Big\vert_{\lambda_i=-aNt_i^2}
\right\}
\\&\quad
+O(e^{-\epsilon N}).
\end{split}
\label{bulk density using F}
\end{equation}

The introduction of  the auxiliary variables $\lambda_1,\lambda_2$ in
(\ref{bulk F definition}) is
a common ruse applied in the saddle point method (see
e.g. Ref. \cite{Wong}) which we now discuss. Suppose that we
construct the Maclaurin expansion in $t_1,t_2$ of $F_{\sigma}(\lambda,t)$ with $\lambda$ considered as a fixed
parameter
\begin{equation}
F_{\sigma}(\lambda,t)=\sum_{j=0}^p\sum_{k=0}^j\frac{t_1^k\, t_2^{j-k}}{k!(j-k)!}
\left[\frac{\partial^k}{ \partial s_1^k}
\frac{\partial^{j-k}}{\partial s_2^{j-k}}
\,F_{\sigma}(\lambda,s_1,s_2)\right]\Big\vert_{s_1,s_2=0}
+O(t_1^{p_1}\,t_2^{p_2})\Big\vert_{p_1+p_2=p+1}
\label{F series}
.
\end{equation}
If we now set $\lambda_i=-aNt_i^2$ in (\ref{F series}) and perform the
integrations required in (\ref{bulk density using F}) then we find that each
term corresponding to a given value of $j$ in (\ref{F series}) has the
same resulting $N$ dependence. This then gives a systematic way of
obtaining the corrections out to any given order in $N$. To see why
this occurs, first note that we need only consider the terms in (\ref{F series})
for which both $j$ and $k$ are even since any odd monomials are
annihilated by (\ref{bulk integral operator}), and then further note that with  $\lambda_i=-aNt_i^2$ we have
% NON-PREPRINT VERSION
\begin{equation}
N\,\mathbb{E}^{(B)}_1\mathbb{E}^{(B)}_2
\,t_1^{2m_1}\,t_2^{2m_2}
\,\lambda_1^{l_1} \lambda_2^{l_2}
=(-1)^{l_1+l_2}\frac{\Gamma(m_1+l_1+1/2)\Gamma(m_2+l_2+1/2)}{4\,\pi^2\,a^{m_1+m_2+1}}
\,N^{-(m_1+m_2)} +O(e^{-a \eta^2\,N}),
\label{bulk Gaussian integrals}
\end{equation}
% PREPRINT VERSION
% \begin{equation}
% \begin{split}
% N\,\mathbb{E}^{(B)}_1\mathbb{E}^{(B)}_2
% \,t_1^{2m_1}\,t_2^{2m_2}
% \,\lambda_1^{l_1} \lambda_2^{l_2}
% &=(-1)^{l_1+l_2}\frac{\Gamma(m_1+l_1+1/2)\Gamma(m_2+l_2+1/2)}{4\,\pi^2\,a^{m_1+m_2+1}}
% \,N^{-(m_1+m_2)} 
% \\&\quad
% +O(e^{-a \eta^2\,N}),
% \label{bulk Gaussian integrals}
% \end{split}
% \end{equation}
for any $l_1,l_2\in\mathbb{N}$. Hence despite the fact that various powers of
$\lambda_1$ and $\lambda_2$ arise when $F_{\sigma}(\lambda,t)$ is differentiated, for a given
value of $j$ all terms end up with the same $N$ dependence after
setting $\lambda_i=-aNt_i^2$ and performing the integrations.

Hence, substituting (\ref{F series}) into (\ref{bulk density using F})
one can construct the asymptotic series for $\rho_N(x)$ out to any
desired order. We shall explicitly construct this series out to order
$1/N$ but the generalization to higher orders is obvious. However, as
we shall see the resulting asymptotic expansions obtained by keeping
only the $1/N$ correction are already extremely good, and it's quite
likely that optimal truncation occurs at this order, providing little
incentive to construct higher order corrections.

Let us denote by $c^{(\sigma)}_{m}(x)$ the coefficient of $1/N^m$ in the $1/N$ expansion 
generated by acting on (\ref{F series}) with 
$N\,\mathbb{E}^{(B)}_1\,\mathbb{E}^{(B)}_2$. Then since the only terms which
contribute are those for which $j$ and $k$ are even
we have 
% NON-PREPRINT VERSION
\begin{equation}
\frac{c^{(\sigma)}_{m}(x)}{N^m}=\sum_{k=0}^m\frac{1}{(2k)!(2[m-k])!}
N\,\mathbb{E}^{(B)}_1\mathbb{E}^{(B)}_2\,t_1^{2k}\,t_2^{2(m-k)} 
\left[\frac{\partial^{2k}}{\partial {s_1}^{2k}}\,\frac{\partial^{2(m-k)}}{\partial {s_2}^{2(m-k)}}
\, F_{\sigma}(\lambda,s)\Bigg\vert_{s_i=0}\right]\Bigg\vert_{\lambda_i=-a\,N\,t_i^2}.
\label{bulk c}
\end{equation}
% PREPRINT VERSION
% \begin{equation}
% \frac{c^{(\sigma)}_{m}(x)}{N^m}=\sum_{k=0}^m\frac{1}{(2k)!(2[m-k])!}
% N\,\mathbb{E}^{(B)}_1\mathbb{E}^{(B)}_2 t_1^{2k} t_2^{2(m-k)} \!
% \left[\frac{\partial^{2k}}{\partial {s_1}^{2k}}\frac{\partial^{2(m-k)}}{\partial {s_2}^{2(m-k)}}
% F_{\sigma}(\lambda,s)\Bigg\vert_{s_i=0}\right]\Bigg\vert_{\lambda_i=-aNt_i^2}.
% \label{bulk c}
% \end{equation}
We can re-express $\rho_N(x)$ from (\ref{bulk density using F})
in terms of the $c^{(\sigma)}_{m}(x)$ as follows

\begin{equation}
\begin{split}
\rho_N(x)&=
2\, h_0\, \re\left\{c^{(1,-1)}_{0}(x)\right\}
-2\, h_0\, \re\left\{e^{2i(\theta-N\,\pi\,P(x))}\,c^{(1,1)}_{0}(x)\right\}
\\&\quad
-2\, h_0\,
\re\left\{e^{2i(\theta-N\,\pi\,P(x))}\,c^{(1,1)}_{1}(x)\right\}
\frac{1}{N}
+
\left[2 h_0\re\left\{c^{(1,-1)}_{1}(x)\right\}+h_1\right]\frac{1}{N}
\\&\quad +O\left(\frac{1}{N^2}\right).
\label{bulk density using c}
\end{split}
\end{equation}

It is not hard to show that the term 
$2\, h_0\, \re\left\{c^{(1,-1)}_{0}(x)\right\}$ is equal to the limiting
density $\rho(x)$ from (\ref{bulk limit}) when $|x|<1$ as required, and also that
$c^{(1,1)}_{0}(x)$ vanishes identically (this is actually obvious
from (\ref{G definition}) and (\ref{bulk G transform})).
The explicit construction of the remaining $c^{(\sigma)}_{m}(x)$ required in (\ref{bulk density using c})
is straightforward and we finally obtain the following.

\begin{prop}
Let $\rho_N(x)$ be as defined in (\ref{density definition}), and let
$x$ be fixed with $|x|<1$. Then as $N\to\infty$ we have the following:

For the GUE
\begin{equation}
\rho_N(x)= \rho(x)-\frac{2\cos[2\,N\,\pi\,P(x)]}{\pi^3\, \rho^2(x)}\frac{1}{N}
+O\left(\frac{1}{N^2}\right),
\label{GUE bulk result}
\end{equation}
while  for the LUE
\begin{equation}
\begin{split}
\rho_N(x)
&= \rho(x) 
-
\left(
\frac{\cos(2\,N\,\pi\,P(x)-\alpha\,\pi\,[1+x\,\rho(x)-P(x)])}{\pi^3\,x^2\, \rho^2(x)}
-\frac{\alpha}{\pi^2\,x\, \rho(x)}
\right)\frac{1}{N}
+ O\left(\frac{1}{N^2}\right),
\\
&= \rho(x) 
-
\left(
\frac{\cos(2\,N\,\pi\,P(x)-2\,\alpha\, \Arccos(\sqrt{x}))}{\pi^3\,x^2\, \rho^2(x)}
-\frac{\alpha}{4(1-x)}
\right)\frac{1}{N}
+ O\left(\frac{1}{N^2}\right),
\label{LUE bulk result}
\end{split}
\end{equation}
where $\rho(x)$ is given in (\ref{bulk limit}) and $P(x)$ is the corresponding probability
distribution function, given explicitly in (\ref{P definition}).
\label{bulk proposition}
\end{prop}
As remarked in the 
Introduction, the result (\ref{GUE bulk result}) was obtained previously in \cite{KalischBraak}
again by steepest descent, but starting from an integral
representation derived by super-symmetric arguments rather than
orthogonal polynomials. 

It is interesting to note that it is the distribution
function $P(x)$ of the limiting density $\rho(x)$ which controls the
large $N$ oscillations in the $1/N$ correction to $\rho_N(x)$. 
Note also that while the non-oscillatory correction vanishes at order $1/N$ for the GUE
leaving only an oscillatory correction at this order, the LUE has both
an oscillatory and a non-oscillatory component to its $1/N$
correction. 

\begin{figure}
\includegraphics{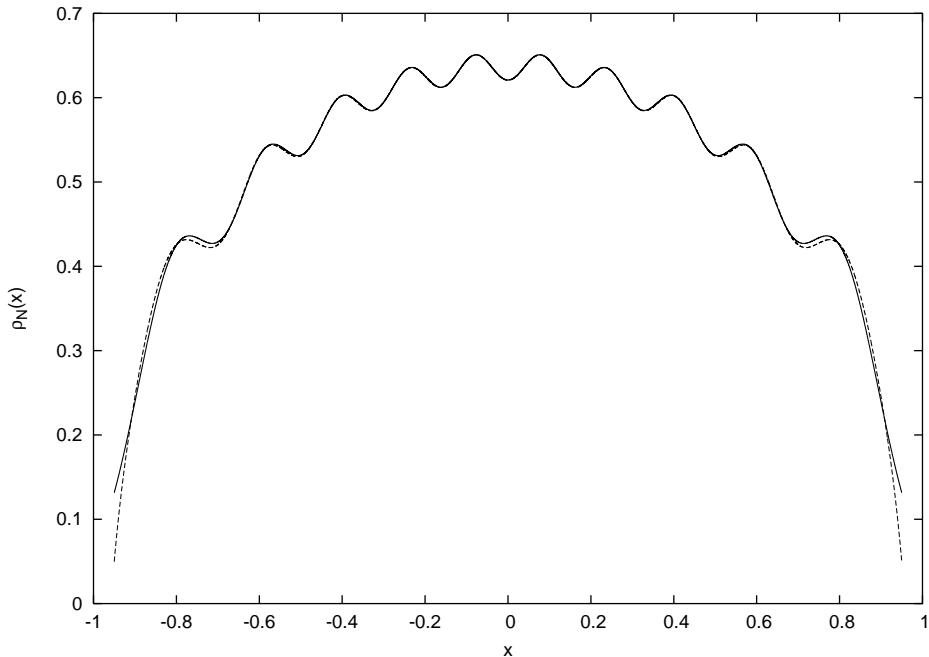}
\caption{\label{bulk gue plot} Comparison of the asymptotic expansion
(\ref{GUE bulk result}), shown as the dashed line, and the exact result
(\ref{Christoffel-Darboux}), shown as the solid line, for the eigenvalue density of the GUE with $N=10$.}
\end{figure}

\begin{figure}
\includegraphics{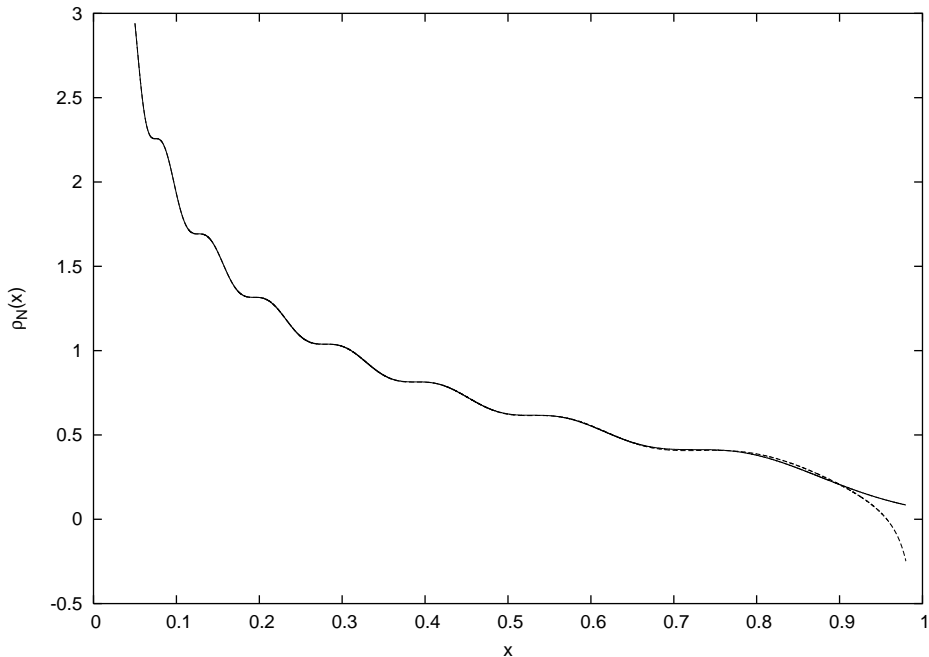}
\caption{\label{bulk lue plot} Comparison of the asymptotic expansion
  (\ref{LUE bulk result}), shown as the dashed line, and the exact result
  (\ref{Christoffel-Darboux}), shown as the solid line, for the eigenvalue density of
the LUE with $\alpha=1/2$ and $N=10$.}
\end{figure}

To demonstrate to the reader just how accurate the expansions given by
Proposition \ref{bulk proposition} are we provide in Figures \ref{bulk gue plot} 
and \ref{bulk lue plot} a numerical comparison of the asymptotic expansions
with the exact results computed using the 
expression (\ref{Christoffel-Darboux}) in terms of orthogonal polynomials .

\section{Soft edge asymptotics for the GUE and LUE}
\label{soft edge}
The appropriate scaling to elucidate the behavior
of $\rho_N(x)$ near the soft edge is to set
$x=1+\xi/N^{2/3}$ for fixed $\xi$, 
as appears in (\ref{soft edge limit}). Substituting such a scaling into (\ref{action definition})
we find
\begin{equation}
N\,S\left(z,1+\frac{b^{1/3}}{2}\frac{\xi}{N^{2/3}}\right)
=
-\xi b^{1/3} N^{1/3} z + N\,S(z),
\label{soft edge action}
\end{equation}
where we've introduced the shorthand $S(z):=S(z,1)$. Here $b>0$ is a
free parameter that we can fix later as convenient. Defining 
\begin{equation}
\jedge:=J_N\left(1+\frac{b^{1/3}}{2}\frac{\xi}{N^{2/3}}\right),
\end{equation}
we see that (\ref{soft edge action})
leads to 
\begin{equation}
\jedge
=
\oint\frac{dz_1}{2\pi i} \exp\left(N\,S(z_1) -b^{1/3}  N^{1/3}\,z_1\, \xi\right)
\oint\frac{dz_2}{2\pi i} \exp\left(N\,S(z_2) -b^{1/3}  N^{1/3}\,z_2\, \xi\right)
G(z_1,z_2).
\label{soft edge start}
\end{equation}
The reader might be concerned by the slightly unorthodox term in the
exponent proportional to $N^{1/3}$, however it is subdominant to the
$N\,S(z)$ term and its presence does not affect any of the usual
arguments of the saddle point method; 
the asymptotic behavior of (\ref{soft edge start}) is determined by
$S(z)$. From (\ref{saddle points}), (\ref{nu definition}) and (\ref{bulk second S derivative}) we see for both the GUE
and LUE that $S(z)$ has one double saddle point, located at $z=-1$. We can
deform the contour of integration to a new contour which 
passes through $z=-1$ along paths of steepest descent. Note
that although in (\ref{pi integrals}) the integrals must be positively
oriented, since we have two integrals in (\ref{soft edge start})
we are free to orient the integrals in the negative direction since the
consequent minus signs cancel. Let us denote by $\mathcal{C}$ the
contour consisting of the union of two rays of unit length, the first
starting at $z=e^{-i\pi/3}-1$ and ending at $z=-1$ and the second
starting at $z=-1$ and ending at $z=e^{i\pi/3}-1$. If we then denote
by $\mathcal{A}$ any suitable arc such that $\mathcal{C}\cup \mathcal{A}$
is a simple closed curve enclosing the origin we can write

\begin{align}
\jedge
&=
\left(\int_{\mathcal{C}}+\int_{\mathcal{A}}\right)
\frac{dz_1}{2\pi i} \exp\left(N\,S(z_1) -b^{1/3}  N^{1/3}\,z_1\,\xi\right)
%\nonumber\\&\quad
\left(\int_{\mathcal{C}}+\int_{\mathcal{A}}\right)
\frac{dz_2}{2\pi i} \exp\left(N\,S(z_2) -b^{1/3}  N^{1/3}\,z_2\, \xi\right)
G(z_1,z_2),
\label{soft edge before throw-away}
\\
&=
\int_{\mathcal{C}}
\frac{dz_1}{2\pi i} \exp\left(N\,S(z_1) -b^{1/3}  N^{1/3}\,z_1\,\xi\right)
%\nonumber\\&\quad
\int_{\mathcal{C}}
\frac{dz_2}{2\pi i} \exp\left(N\,S(z_2) -b^{1/3}  N^{1/3}\,z_2\, \xi\right)
G(z_1,z_2)
%\nonumber\\&\quad
+o(e^{2N\,S(-1)} N^{-p}).
\label{soft edge on C}
\end{align}
The error bound in (\ref{soft edge on C}) holds for all
$p\in\mathbb{N}$, and so in what follows we consider $p$ as
arbitrarily large.
The equality between (\ref{soft edge on C}) and 
(\ref{soft edge before throw-away}) can be obtained by noting that we can choose $\mathcal{A}$
to consist of two rays lying along the path of steepest descent away
from the endpoints of $\mathcal{C}$, which extend as far as we like
into the right half plane, together with an arc to close the contour which
we can choose to be as far into the right half plane as desired. With
such a choice for $\mathcal{A}$ one can obtain the required bounds by a
straightforward generalization of the usual argument used in the saddle point method.
For a careful discussion of the saddle point method 
suitable for this purpose
see for example Section 2.5 of Ref. \cite{Wong}. We note that
Ref. \cite{Wong} refers to the saddle point method as {\em Perron's method}.

Now let us change variables in (\ref{soft edge on C}) according to
$t=z+1$, so that the vertex of our contour is now at the origin.
We shall denote the image of $\mathcal{C}$ under this change of
variables by $\mathcal{B}$. Further, since $S(z)$ is analytic on
$\mathcal{C}$ we have
\begin{equation}
S(t-1)=S(-1)+b \frac{z^3}{3} +b \frac{z^3}{3}\,\varphi(t),
\end{equation}
where 
\begin{equation}
\varphi(t):=\sum_{k=4}^{\infty}\frac{S^{(k)}(-1)}{S^{(3)}(-1)}\frac{3!}{k!}\,t^{k-3},
\end{equation}
and we have now chosen 
\begin{align}
b
&
=\frac{S^{(3)}(-1)}{2},
\\
&=
\begin{cases}
1,&\gue,\\
2,&\lue.\\
\end{cases}
\end{align}
We also note that by setting $x=1$ in (\ref{Re S}) and (\ref{Im S}) we have
\begin{equation}
S(-1)=
\begin{cases}
\displaystyle
\frac{3}{2}-i\pi,&\gue,\\
2-\log(2)-i\pi,&\lue.\\
\end{cases}
\end{equation}
Finally, defining
\begin{equation}
F(\lambda,t):=
G(t_1-1,t_2-1)\,
e^{\lambda_1\,\varphi(t_1)+\lambda_2\,\varphi(t_2)},
\end{equation}
and 
\begin{equation}
\mathbb{E}^{(S)}_i:=\int_{\mathcal{B}}
\exp\left(b N \frac{t_i^3}{3}-\xi b^{1/3} N^{1/3}t_i \right)\frac{dt_i}{2\pi i}
\end{equation}
we have
\begin{equation}
\jedge=e^{2N \re[S(-1)]+ 2b^{1/3} N^{1/3}\,\xi}
\left[
\,\mathbb{E}^{(S)}_1\,\mathbb{E}^{(S)}_2\,
F(\lambda,t)\vert_{\lambda_i=b N t_i^3/3}+
o(N^{-p})\right].
\label{last soft J}
\end{equation}
If we now multiply (\ref{last soft J}) by the prefactors required in 
(\ref{density double integral}) using (\ref{prefactor asymptotics}) we obtain
\begin{equation}
\begin{split}
\rhosoft
&=
\left[g_0(\xi)+\frac{g_1(\xi)}{N^{1/3}}
+\frac{g_2(\xi)}{N^{2/3}}
+O\left(\frac{1}{N}\right)
\right]
%\\&\quad\times
\left[
N^{4/3}\,\mathbb{E}^{(S)}_1\mathbb{E}^{(S)}_2
F(\lambda,t)\Big\vert_{\lambda_i=b N t_i^3/3}
+o(N^{-p})
\right],
\end{split}
\label{soft edge density with g}
\end{equation}
where $g_m(\xi)$ is the coefficient of $N^{-m/3}$ in the large $N$
fixed $\xi$ expansion of 
\begin{equation}
\begin{cases}
\displaystyle
e^{-\xi^2/2N^{1/3}},&\gue,\\
\displaystyle
2^{2\alpha-2/3}\left(1+\frac{\xi}{(2N)^{2/3}}\right)^{\alpha},&\lue.\\
\end{cases}
\end{equation}
In (\ref{soft edge density with g}) we have presented only terms $O(1/N)$ in the first factor since this is
will be sufficient for our purposes in what follows. Higher order terms are easily
retained if desired.

Our work is now essentially done. One expands $F(\lambda,t)$
around $t=0$ for fixed $\lambda$ as in (\ref{F series}) and then sets $\lambda_i=b N t_i^3/3$,
analogous to the bulk case. Again, after integration, each value of $j$ in the Maclaurin expansion
(\ref{F series}) contributes to the same order in $N$. To see this explicitly
we can use the following lemma.

\begin{lemma}
Let $\mathcal{B}$ be the contour consisting of the union of a ray
starting at $e^{-i \pi/3}$ and ending at the origin, and a ray
starting at the origin and ending at $e^{i\pi/3}$. For any
$b>0$ and $0<\beta<1/3$ we have for large $N$ that  
$$
\int_{\mathcal{B}}z^m\,\exp\left(b N \frac{z^3}{3}-\xi b^{1/3} N^{1/3}
z\right)\frac{dz}{2\pi i}
=
(-1)^m
b^{-(m+1)/3}\,N^{-(m+1)/3}[\ai^{(m)}(\xi)+ O(e^{-\beta b N})],$$
where $\ai^{(m)}(\xi)$ is the $m^{th}$ derivative of the Airy function $\ai(\xi)$.
\label{airy integration}
\end{lemma}
\begin{proof}
This follows from the standard entire contour integral expression for
$\ai(\xi)$ (see e.g. \cite{Olver}) by simply changing
variables $z\mapsto b^{-1/3} N^{-1/3}z$, and noting that rays defining
the contour $\mathcal{B}$
can be extended to infinity at the cost of introducing exponentially
subdominant corrections.
\end{proof}
An immediate consequence of Lemma \ref{airy integration} is that with
$\lambda_i=bNt_i^3/3$ we have
% NON-PREPRINT VERSION
\begin{equation}
N^{2/3}\,\mathbb{E}^{(S)}_1\,\mathbb{E}^{(S)}_2\,t_1^{m_1}\,t_2^{m_2}\,\lambda_1^{l_1}\,\lambda_2^{l_2}
=
\frac{(-1)^{l_1+l_2+m_1+m_2}}{3^{l_1+l_2}\,b^{(m_1+m_2+2)/3}}
\ai^{(m_1+3l_1)}(\xi)\,\ai^{(m_2+3l_2)}(\xi)\,\left(\frac{1}{N^{1/3}}\right)^{m_1+m_2}
+O(e^{-\beta b N}).
\label{soft edge integration lemma}
\end{equation}
% PREPRINT VERSION
% \begin{equation}
% \begin{split}
% N^{4/3}\,\mathbb{E}^{(S)}_1\,\mathbb{E}^{(S)}_2\,t_1^{k_1}\,t_2^{k_2}\,\lambda_1^{l_1}\,\lambda_2^{l_2}
% &=
% \frac{(-1)^{l_1+l_2+k_1+k_2}}{3^{l_1+l_2}\,b^{(k_1+k_2+2)/3}}
% \ai^{(k_1+3l_1)}(\xi)\,\ai^{(k_2+3l_2)}(\xi)\,\left(\frac{1}{N^{1/3}}\right)^{k_1+k_2-2}
% \\&\quad+O(e^{-\beta b N}).
% \end{split}
% \label{soft edge integration lemma}
% \end{equation}
Hence, if we construct the Maclaurin expansion of $F(\lambda,t)$ with
$\lambda$ fixed as in (\ref{F series}), and set
$\lambda_i=bNt_i^3/3$ and integrate using (\ref{soft edge integration lemma}),
we obtain an expansion for 
\begin{equation}
N^{4/3}\,\mathbb{E}^{(S)}_1\,\mathbb{E}^{(S)}_2\,F(\lambda,t)\vert_{\lambda_i=b N t_i^3/3}
\label{soft edge F series}
\end{equation}
in powers of $N^{-1/3}$. Denoting  the coefficient of $N^{-m/3}$ in
this expansion by $c_m(\xi)$ we have explicitly that 
\begin{equation}
c_m(\xi)=\sum_{k=0}^{m+2}\!
\frac{1}{k!(m+2-k)!}N^{4/3}\mathbb{E}^{(S)}_1\mathbb{E}^{(S)}_2 t_1^{k}\,t_2^{m+2-k}
\!\! \left[\frac{\partial^k}{ \partial s_1^k}
 \frac{\partial^{m+2-k}}{\partial s_2^{m+2-k}}
 \,F(\lambda,s_1,s_2)\Big\vert_{s_1,s_2=0}\right]
\! \Big\vert_{\lambda_i=b N t_i^3/3}.
\label{soft edge c} 
\end{equation}
The reader might be concerned that
according to (\ref{soft edge integration lemma}) the $k_1+k_2=0$
and $k_1+k_2=1$ terms grow with $N$; however
it is not hard to show from (\ref{soft edge c}) that the coefficients $c^{(-2)}(\xi)$ and $c^{(-1)}(\xi)$
vanish identically. 

\begin{figure}
\includegraphics{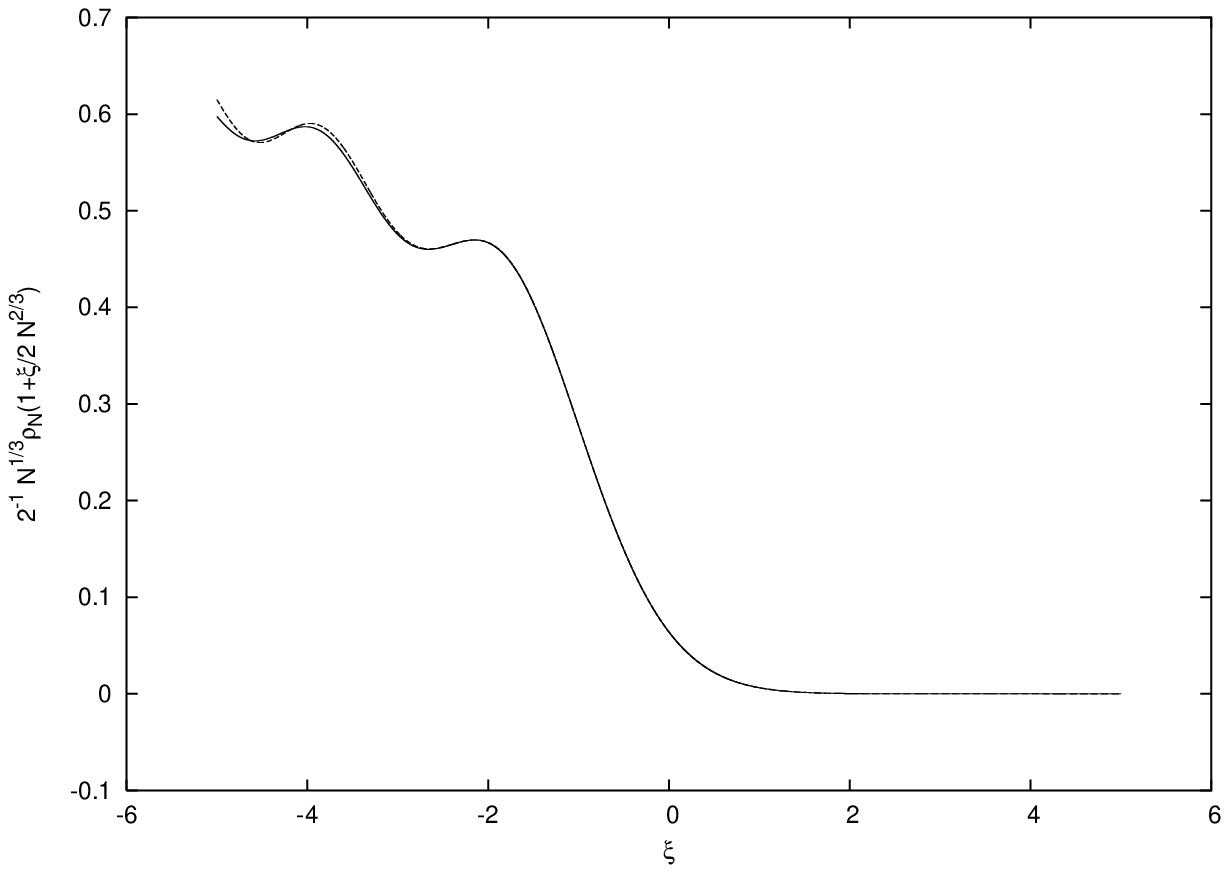}
\caption{\label{edge gue plot} Comparison of the asymptotic expansion
  (\ref{GUE edge result}), shown as the dashed line, and the exact result
  (\ref{Christoffel-Darboux}), shown as the solid line, for the eigenvalue density near the
  soft edge at $x=1$, for the GUE with $N=10$.}
\end{figure}

\begin{figure}
\includegraphics{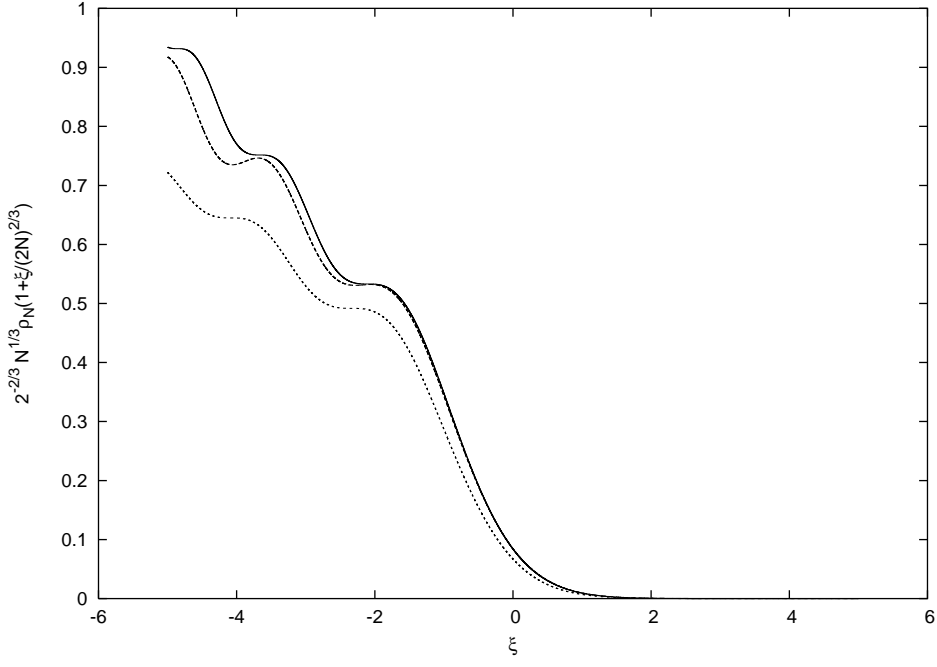}
\caption{\label{edge lue plot}Comparison of the asymptotic expansion
  (\ref{LUE edge result}), shown as the dashed line, and the exact result
  (\ref{Christoffel-Darboux}), shown as the solid line, for the eigenvalue density near the
  soft edge at $x=1$, for the LUE with $\alpha=1/2$ and $N=20$. Also
  shown is the limit as $N\to\infty$ given by the Airy kernel
  (\ref{soft edge limit}), shown as the dotted line lying below the
  other two curves.}
\end{figure}

We can now express (\ref{soft edge density with g}) in terms of the
coefficients $c_m(\xi)$ as 
\begin{equation}
\begin{split}
\rhosoft
&=
g_0(\xi)\,c_0(\xi)+[g_1(\xi)\,c_0(\xi)+g_0(\xi)\,c_1(\xi)]\frac{1}{N^{1/3}}
\\
&\quad+ [g_2(\xi)\,c_0(\xi)+g_1(\xi)\,c_1(\xi)+g_0(\xi)\,c_2(\xi)]\frac{1}{N^{2/3}}
+O\left(\frac{1}{N}\right).
\end{split}
\label{final soft edge density}
\end{equation}
We have explicitly displayed terms
$o(1/N)$ here, but it straightforward to retain as many
terms as desired. The expansion constructed from terms $o(1/N)$
however is extremely accurate, as we demonstrate in Figures 
\ref{edge gue plot} and \ref{edge lue plot}, and it appears to be the
numerically optimal order
at which to truncate the expansions.

The explicit forms for the coefficients $c_m(\xi)$ can be constructed 
from (\ref{soft edge c}) and
substituted into (\ref{final soft edge density}). We can also further simplify the Airy
derivatives appearing in (\ref{soft edge integration lemma}) 
using the Airy differential equation $\ai''(\xi)=\xi\,\ai(\xi)$
so that only $\ai(\xi)$ and its first derivative appear. We finally
obtain the following.

\begin{prop}
Let $\rho_N(x)$ be as defined in (\ref{density definition}). 
Then with $\xi$ fixed, as $N\to\infty$ we have the following:

For the GUE
\begin{equation}\begin{split}
\frac{N^{1/3}}{2}\rho_N\left(1+\frac{\xi}{2 N^{2/3}}\right)&=
[\aip(\xi)]^2-\xi [\ai(\xi)]^2
\\
&\quad -\frac{1}{20}\left(3\xi^2[\ai(\xi)]^2 -2\xi[\aip(\xi)]^2 -3\ai(\xi)\aip(\xi)\right)\frac{1}{N^{2/3}}
\\& \quad
+O\left(\frac{1}{N}\right),
\end{split}
\label{GUE edge result}
\end{equation}
while for the LUE
\begin{equation}\begin{split}
\frac{(2N)^{1/3}}{2}\rho_N\left(1+\frac{\xi}{(2N)^{2/3}}\right)& =
[\aip(\xi)]^2-\xi[\ai(\xi)]^2
+ \frac{\alpha}{2^{1/3}} [\ai(\xi)]^2  \frac{1}{N^{1/3}}
\\& \quad
+\!\frac{2^{1/3}}{10}\!\left(3\xi^2[\ai(\xi)]^2-2\xi[\aip(\xi)]^2+(2-5\alpha^2)\ai(\xi)\aip(\xi) \right)\frac{1}{N^{2/3}}
\\& \quad
+O\left(\frac{1}{N}\right).
\end{split}
\label{LUE edge result}
\end{equation}
\label{edge proposition}
\end{prop}

Figures \ref{edge gue plot} 
and \ref{edge lue plot} provide a numerical comparison of the asymptotic expansions
given in Proposition \ref{edge proposition} with the exact results computed using the 
expression (\ref{Christoffel-Darboux}) in terms of orthogonal polynomials.

Note that it appears that the GUE converges much faster than the
LUE, since while the two curves in (\ref{edge gue plot}) are almost
indistinguishable at $N=10$ for the GUE, the asymptotic expansion for
the LUE begins to diverge from the exact result in (\ref{edge lue
  plot}) already by $\xi \sim 2$, and both curves are rather different
from the limiting Airy kernel expression (\ref{soft edge limit}). We
investigated the affects of retaining more terms in the expansion for
the LUE case; keeping terms $O(1/N)$ did not noticeably change the
plots, while keeping terms higher than $1/N$ caused significant
divergence of the asymptotic expansion from the exact result. 
The explanation for this is most likely that,
as made precise in the next section, the edge
expansions match onto the bulk expansions,
and these in turn become more divergent near the
edge at each order in $1/N$.

\section{Matching of the bulk and edge expansions}
\label{matching}
In Figures \ref{bulk gue plot}--\ref{edge lue plot} plots of the bulk and edge asymptotic
expansions have separately been compared against the exact density for $N = 10$. Although the
scale of the independent variable is different, we can see from Figures \ref{edge gue plot}
and \ref{edge lue plot} that the edge asymptotic expansions are accurate approximations to the
exact density up to the neighborhood of the first local maximum (relative to the edge $\xi = 0$)
at least and thus should be used instead of the bulk asymptotic expansion in this region.

At a quantitative level, it is possible to exhibit a matching between the various asymptotic
expansions. Suppose in (\ref{GUE bulk result}) we set $x=1 + \xi/2N^{2/3}$, and in
(\ref{LUE bulk result}) we set $x=1 + \xi/(2N)^{2/3}$, and take $\xi < 0$ and fixed.
Expanding the right hand sides as an asymptotic series in $N$ gives
\begin{align}
N^{1/3} 
\rho_N^{\rm GUE}(1 + \xi/2 N^{2/3})
&\asym
\left(\frac{2 \sqrt{|\xi|}}{\pi} - 
\frac{\cos(4 |\xi|^{3/2}/3)}{2 \pi |\xi|} \right) 
\nonumber\\& \quad 
- 
\Big ( \frac{|\xi|^{3/2}}{4 \pi} +
\frac{\cos(4 |\xi|^{3/2}/3)}{ 8 \pi} +
\frac{|\xi|^{3/2} \sin (4 |\xi|^{3/2}/3)}{20 \pi} \Big )
\frac{1}{N^{2/3}} 
\nonumber\\&\quad
+O \left( \frac{1}{N^{4/3}} \right).
\label{gue matching}
\\
(2N)^{1/3} 
 \rho_N^{\rm LUE}(1 + \xi/(2 N)^{2/3})
&\asym
\left( \frac{2 \sqrt{|\xi|}}{\pi} - 
 \frac{\cos(4 |\xi|^{3/2}/3)}{2 \pi |\xi|} \right)
\nonumber\\& \quad
 + \frac{\alpha (1 + \sin(4 |\xi|^{3/2}/3))}{\pi \sqrt{|\xi|}} 
\frac{1}{(2N)^{1/3} } 
+
O\left( \frac{1}{N^{2/3}}\right)
\label{lue matching}
\end{align}
where the symbol 
$\mathop{\sim}\limits^{.}$ 
denotes that the asymptotic series have been
expanded as specified. An important feature is that this procedure mixes the terms which are
at different orders in $N$ in (\ref{GUE bulk result}) and (\ref{LUE bulk result}).

Let us now compute the $\xi \to - \infty$ asymptotic expansions of the right hand sides of the
first two terms in each of (\ref{GUE edge result}) and (\ref{LUE edge
  result}), multiplied by $2$.  Using the fact
that for $x \to \infty$ (see e.g.~\cite{Olver})
$$
\ai(-x) \: \sim \: \frac{1}{\sqrt{\pi} x^{1/4} }
\cos\Big ( \frac{\pi}{4} - \frac{2}{3} x^{3/2} \Big ) -
\frac{5 \sqrt{3}}{24 x^{1/2} } \cos \Big ( \frac{\pi}{4} + \frac{2}{3} x^{3/2} \Big ) +
O \Big ( \frac{1}{x^{3/4}} \Big )
$$
we obtain expansions which reproduce the $N$-independent terms in (\ref{gue matching}) and (\ref{lue matching}),
giving furthermore, terms of higher order in $1/|\xi|$. In (\ref{lue matching}) the term proportional to
$1/(2N)^{1/3}$ is reproduced, and this too is accompanied by terms of higher order in
$1/|\xi|$. In (\ref{gue matching}) the term $-|\xi|^{3/2}/4 \pi N^{2/3}$ is reproduced, while the other
terms proportional to $1/N^{2/3}$ are out by rational factors. The explanation for the missing
higher order terms in $1/|\xi|$, and incorrect rational factors is most likely due to the fact
that terms of all orders in $1/N$ in (\ref{GUE bulk result}) and (\ref{LUE bulk result})
contribute to each distinct order in the expansions (\ref{gue matching}) and (\ref{lue matching}).
Specifically, from the results exhibited above, it would seem that expanding the complete
large $N$ asymptotic series for $\rho_N^{\rm GUE}(x)$ and $\rho_N^{\rm LUE}(x)$ as in
(\ref{gue matching}) and (\ref{lue matching}) would give precisely the large $\xi \to - \infty$ expansion of
 (\ref{GUE edge result}) and (\ref{LUE edge result}), extended to all orders in $N$.

% \bibliography{bib}

\end{document}